\documentclass[12pt]{article}

\begin{document}

\newcommand{\be}{\begin{equation}}
\newcommand{\ee}{\end{equation}}
\newcommand{\bea}{\begin{eqnarray}}
\newcommand{\eea}{\end{eqnarray}}
\newcommand{\ack}[1]{[{\bf Pfft!: #1}]}

\newcommand{\eref}[1]{eq.\ (\ref{eq:#1})}
\def\NPB{{\it Nucl. Phys. }{\bf B}}
\def\PL{{\it Phys. Lett. }}
\def\PRL{{\it Phys. Rev. Lett. }}
\def\PRD{{\it Phys. Rev. }{\bf D}}
\def\CQG{{\it Class. Quantum Grav. }}
\def\JMP{{\it J. Math. Phys. }}
\def\SJNP{{\it Sov. J. Nucl. Phys. }}
\def\SPJ{{\it Sov. Phys. J. }}
\def\JETPL{{\it JETP Lett. }}
\def\TMP{{\it Theor. Math. Phys. }}
\def\IJMPA{{\it Int. J. Mod. Phys. }{\bf A}}
\def\MPL{{\it Mod. Phys. Lett. }}
\def\CMP{{\it Commun. Math. Phys. }}
\def\AP{{\it Ann. Phys. }}
\def\PR{{\it Phys. Rep. }}

\hyphenation{Min-kow-ski}
\hyphenation{cosmo-logical}
\hyphenation{holo-graphy}
\hyphenation{super-symmetry}
\hyphenation{super-symmetric}

\centerline{\Large \bf }\vskip0.25cm
\centerline{\Large \bf }\vskip0.25cm
\centerline{\Large \bf Transient Astrophysical Pulses and Quantum Gravity}\vskip0.25cm
\vskip 1cm

\renewcommand{\thefootnote}{\fnsymbol{footnote}}
\centerline{{\bf Michael Kavic\footnote{kavic@vt.edu},
Djordje Minic\footnote{dminic@vt.edu},
and John Simonetti\footnote{jhs@vt.edu}}}
\vskip .5cm
\centerline{\it Institute for Particle, Nuclear and Astronomical Sciences}
\centerline{\it Department of Physics, Virginia Tech}
\centerline{\it Blacksburg, VA 24061, U.S.A.}


\vskip .5cm

\begin{abstract}
Searches for transient astrophysical pulses could open an exciting new window into the
fundamental physics of quantum gravity. In particular, an evaporating primordial black hole
in the presence of an extra dimension can produce a detectable transient pulse.
Observations of such a phenomenon can in principle explore the electroweak energy scale, indicating that astrophysical
probes of quantum gravity can successfully complement the exciting new physics expected to be discovered
in the near future at the Large Hadron Collider.

\end{abstract}

\setcounter{footnote}{0}
\renewcommand{\thefootnote}{\arabic{footnote}}

\newpage

\section{Introduction}

For millennia, astronomy has concentrated on the unchanging nature of the cosmos.  
Even throughout the 20th century, astrophysical theory and observation focused on 
the physics of persistent objects --- stars, galaxies, etc.  Observations were often directed at a single target, and for as long as possible to obtain high 
precision measurements. However, high energy events may occur in seemingly 
random parts of the sky, over a short time scale, and could be missed 
if traditional astronomical methods are employed. It is our contention that 
such transients are just the type of phenomena that could be related to quantum 
gravitational effects, and that searches for these could provide a new arena in which to probe
this elusive area of inquiry. 

A theory of quantum gravitation would deepen our understanding of spacetime, matter,
and the origin of the universe \cite{Jejjala:2007rn, Jejjala:2007hh}.
However, it is crucial that any such theory be subject to experimental and observational verification. Observable effects of quantum gravity are expected to manifest themselves most directly 
at exceptionally high energies or in the presence of a spacetime singularity. 
Thus, experimental tests of quantum gravity present a severe challenge. Particle accelerators 
have long served as an indispensable tool for exploring new regimes of fundamental physics, but it may be some time before they yield a discernible signature of quantum gravitation. 
Given the extreme difficulties posed by the search for quantum gravitational effects,  
another source of data would be of great value, and could provide a means of comparison
with accelerator-based experiments. 

\section{Transient Pulses from Exploding Primordial Black Holes}

Here we consider just one example of a transient event, associated with the explosion of primordial 
black holes (PBHs), which depends upon two distinct quantum gravitational phenomena: Hawking radiation and 
the existence of an extra spatial dimension.

The defining relation governing the Hawking evaporation of a black hole
\cite{Hawking} is 
\be
T = \frac{\hbar c^3}{8\pi G k} \frac{1}{M},
\label{tem}
\ee
for mass $M$ and temperature $T$. The power emitted by the black hole is
\be
P \propto \frac{\alpha(T)}{M^2},
\label{pow}
\ee
where $\alpha(T)$ is the number of particle modes available.
Equations~(\ref{tem}) and (\ref{pow}), along with an increase in the
number of particle modes available at high temperature, leads to the
possibility of an explosive outburst as the black hole evaporates its
remaining mass in an emission of radiation and particles. PBHs of sufficiently
low mass would be reaching this late stage now \cite{Carr,Khlopov}.
Searches for these explosive outbursts have traditionally focused on
$\gamma$-ray detection \cite{Halzen}. However, Rees noted that exploding
primordial black holes could provide an observable coherent radio pulse
that would be easier to detect \cite{Rees}.

Rees and Blandford \cite{Rees,Blandford} describe the production of a
coherent electromagnetic pulse by an explosive event in which the entire
mass of the black hole is emitted. If significant numbers of
electron-positron pairs are produced in the event, the relativistically
expanding shell of these particles (a ``fireball'' of Lorentz factor
$\gamma_f$) acts as a perfect conductor, reflecting and boosting the
virtual photons of the interstellar magnetic field. An electromagnetic
pulse results only for $\gamma_f \sim 10^5$ to $10^7$, for typical
interstellar magnetic flux densities and free electron densities. The
energy of the electron-positron pairs is
\begin{equation}
\label{gen1}
kT \approx \frac{\gamma_f}{10^5} 0.1 \ {\rm TeV}. 
\end{equation}
Thus the energy associated with $\gamma_f\sim10^5$ corresponds roughly
to the electroweak scale.

\section{Exploding Primordial Black Holes \& the TeV Scale}

There exists a remarkable relationship between the
range of pulse-producing Lorentz factors for the emitted particles, and
the TeV scale \cite{Kavic:2008qb}. Since $\gamma_f \propto T$ at the time of the explosive
burst, equation~(1) yields
\begin{equation}
\frac{\gamma_f}{10^5} \approx \frac{10^{-19}\ \textrm{m}}{R_s},
\label{gen2}
\end{equation}
where $R_s$ is the Schwarzschild radius. Thus, the allowed range of Lorentz factors implies 
length scales $R_s\sim 10^{-19} - 10^{-21}$~m.
Taking these as 
Compton wavelengths we find the associated energy scales to be
\begin{equation}
(R_s/\hbar c)^{-1}\sim 1-100 \ \textrm{TeV}.
\label{generic2}
\end{equation}
This relationship suggests that the production of an electromagnetic
pulse by PBHs might be used to probe TeV-scale physics. 

On general grounds one could expect quantum gravity to
probe other scales that are vastly different from the  experimentally forbidding
Planck energy region.
The effective action for gravity coupled to Standard Model matter carries additional 
information apart from a sensitivity to the Planck scale.
It also encodes information about the scale associated with the
vacuum energy density, which also sets the relevant cosmological scale, 
as described by the cosmological constant, as well as the energy scale relevant for 
the generation of particle masses.
In the case of the Standard Model of particle physics this corresponds to 
the electroweak scale. Also, by unitarity, new phenomena are expected at
a characteristic scale of the order of a TeV.
Thus PBH explosions, and therefore the accompanying quantum gravitational 
phenomena, can be sensitive to 
other energy scales,
such as the electroweak scale or the TeV scale, which are many orders of magnitude different from
the Planck scale.

\section{Transient Pulses from Primordial Black Holes in the Presence
of an Extra Dimension}

To make use of
these interesting generic observations, a specific
phenomenologically relevant explosive process is required. One such
process, discussed by Kol \cite{Kol1}, which connects quantum gravitational phenomena and the TeV
scale, makes use of the possible existence of an extra dimension and relies on the physics of
the black string/black hole phase transition.  

Black holes in four dimensions are uniquely defined by charge, mass, and angular momentum.  
However, with the addition of an extra spatial dimension, black holes could exist in different 
phases and undergo phase transitions.  For one toroidally compactified extra dimension of length $L$, 
two possible phases are a black string wrapping the compactified extra dimension, and a 5-dimensional black hole 
smaller than the extra dimension. A topological phase transition from the black string to the black hole occurs 
when an instability, the Gregory-LaFlamme point \cite{GL}, is reached.  This transition is of 
first order \cite{Gubser}, and results in a significant release of energy equivalent to a substantial increase 
in the luminosity of Hawking radiation \cite{Kol2}.
The
sensitivity to widely separate energy scales 
can be nicely encapsulated in one overall efficiency parameter $\eta$ characterizing
the PBH explosion.

The analysis of Rees and Blandford \cite{Rees,Blandford} can be adapted to the topological phase 
transition scenario.
Frequencies between $\sim$~1~GHz and $10^{15}$~Hz ($\gamma_f \sim 10^5$ to $10^7$) 
sample possible extra dimensions 
between $L\sim 10^{-18}-10^{-20}$~m.
These length scales correspond to energies of $(L/\hbar c)^{-1}\sim0.1-10$~TeV. 
The efficiency parameter and the expected linear polarization of the pulse will 
make this transient distinguishable 
from other possible sources \cite{Kavic:2008qb}. 

The electroweak scale 
is $\sim$0.1~TeV, and thus, radio observations at $\sim$~1~GHz may be most significant. 
Radio observations have historically been significant as probes of new physics, for example in the 
discovery and mapping of the Cosmic Microwave Background. Observations of transient radio phenomena have also played 
a role in astrophysical exploration, in the discovery of pulsars. 
A new generation of radio telescopes will search for transient radio pulses
\cite{Bower,LWA,MWA,LOFAR,ETA1,ETA2}. Such searches,
using pre-existing data, have recently found surprising pulses of
galactic and extragalactic origin \cite{McLaughlin,Lorimer,Vachaspati}.
The Eight-meter-wavelength Transient Array (ETA) \cite{ETA1,ETA2}, for example, can detect
the type of pulse discussed here, out to a distance of order 100~pc.

In the case of TeV-scale compactification models in which both gauge
fields and fermions propagate in the extra dimension \cite{UED} the
current bound is $(L/\pi\hbar c)^{-1}\sim300-500$~GeV with the 
Large Hadron Collider (LHC)
probing to $\sim 1.5$~TeV \cite{PDG}. Detection of a transient pulse
would imply, as noted above, an extra dimension with $L\sim10^{-18} -
10^{-20}$~m, corresponding to an energy of $\sim 0.1 - 10$~TeV. Thus
constructive comparison of the pulse detection results and LHC results
would be possible.

In the context of the braneworld scenario proposed by Randall and
Sundrum \cite{RS1, RS2} it has been argued that evaporating black holes
will reach a Gregory-Laflamme instability as the radius of the black
hole approaches the AdS radius \cite{champ, emp}. More specifically, in
the Randall-Sundrum I scenario a nominal value of this radius is
10~{TeV}$^{-1}$ \cite{CIN} placing it within the appropriate range for
transient pulse production.

For large extra dimension models \cite{ADD} the effective fundamental
energy scale is much higher than the energy scale of the large extra
dimension $(L/\hbar c)^{-1}$. For a single large extra dimension of size
$L\sim10^{-18} - 10^{-20}$~m the effective fundamental energy scale is
$\sim 10^{10}$~TeV --- much higher than the electroweak scale. Thus,
searches for pulses from topological phase transitions would probe, for
these models, energies inaccessible to accelerator-based approaches for
the foreseeable future.

A positive pulse detection would signal the existence of an extra
dimension, and thus indicate the sensitivity of quantum gravity to scales far below
the Planck scale.  A null detection would serve to constrain the possible size
of an extra dimension, and the relevant low energy scale in particular models. 

%

\section{Discussion \& Outlook}

We have considered only one among a number of possible distinct 
transient events which could reveal new physics. 
The analysis considered here can potentially be extended to  stellar-mass black holes, regardless of their origin, when quantum 
gravitational effects are taken into account, as discussed
in \cite{Emparan:2002jp, Emparan:2002px}.  Another candidate for producing an observable 
transient pulse, which is intimately dependent on quantum gravitational effects, 
is the spark from a cusp of a superconducting cosmic string 
\cite{Vachaspati}. 
Given the connection between highly energetic astrophysical events and 
the production of transient pulses, it is likely 
that searches for these signals will open a new
observational avenue to the heart of quantum gravity.

\vskip .5cm

{\bf \Large Acknowledgements}

\vskip .5cm

We would
like to profoundly thank other members of the ETA collaboration, S. Cutchin, S. Ellingson and C. Patterson, for collaboration regarding the material presented in this  
essay. We also thank G. Djorgovski and N. Kaloper for inspiring communications.  This work is supported by NSF grant AST-0504677,                             and by the Pisgah Astronomical Research Institute. {\small DM}
is supported in part by the U.S. Department of Energy under contract DE-FG05-92ER40677.

\vskip 1cm

\bibliographystyle{unsrt}
\bibliography{references}

\begin{thebibliography}{10}

\bibitem{Jejjala:2007rn}
Vishnu Jejjala, Michael Kavic, and Djordje Minic.
\newblock {Time and M-theory}.
\newblock {\em Int. J. Mod. Phys.}, A22:3317--3405, 2007.

\bibitem{Jejjala:2007hh}
Vishnu Jejjala, Michael Kavic, and Djordje Minic.
\newblock {Fine Structure of Dark Energy and New Physics}.
\newblock {\em Adv. High Energy Phys.}, 2007:21586, 2007.

\bibitem{Hawking}
S.~W. Hawking.
\newblock Particle creation by black holes.
\newblock {\em Commun. Math. Phys.}, 43:199--220, 1975.

\bibitem{Carr}
Bernard~J. Carr.
\newblock Primordial black holes as a probe of cosmology and high energy
  physics.
\newblock {\em Lect. Notes Phys.}, 631:301--321, 2003.

\bibitem{Khlopov}
M.~Yu. Khlopov.
\newblock {Primordial Black Holes}.
\newblock arXiv:0801.0116 [astro-ph], 2008.

\bibitem{Halzen}
F.~{Halzen}, E.~{Zas}, J.~H. {MacGibbon}, and T.~C. {Weekes}.
\newblock {Gamma rays and energetic particles from primordial black holes}.
\newblock {\em Nature}, 353:807--815, October 1991.

\bibitem{Rees}
M.~J. Rees.
\newblock A better way of searching for black-hole explosions?
\newblock {\em Nature}, 266:333--334, 1977.

\bibitem{Blandford}
R.~D. Blandford.
\newblock Spectrum of a radio pulse from an exploding black hole.
\newblock {\em Mon. Not. R. Astron. Soc.}, 181:489--498, 1977.

\bibitem{Kavic:2008qb}
Michael Kavic, John~H. Simonetti, Sean~E. Cutchin, Steven~W. Ellingson, and
  Cameron~D. Patterson.
\newblock {Transient Pulses from Exploding Primordial Black Holes as a
  Signature of an Extra Dimension}.
\newblock arXiv:0801.4023 [astro-ph], 2008.

\bibitem{Kol1}
Barak Kol.
\newblock Explosive black hole fission and fusion in large extra dimensions.
\newblock hep-ph/0207037, 2002.

\bibitem{GL}
R.~Gregory and R.~Laflamme.
\newblock Black strings and p-branes are unstable.
\newblock {\em Phys. Rev. Lett.}, 70:2837--2840, 1993.

\bibitem{Gubser}
Steven~S. Gubser.
\newblock On non-uniform black branes.
\newblock {\em Class. Quant. Grav.}, 19:4825--4844, 2002.

\bibitem{Kol2}
Barak Kol.
\newblock The phase transition between caged black holes and black strings: A
  review.
\newblock {\em Phys. Rept.}, 422:119--165, 2006.

\bibitem{Bower}
Geoffrey~C. Bower.
\newblock Astronomy: Mining for the ephemeral.
\newblock {\em Science}, 318:759--760, 2007.

\bibitem{LWA}
G.~B. {Taylor}.
\newblock {The Long Wavelength Array}.
\newblock {\em Long Wavelength Astrophysics, 26th meeting of the IAU, Joint
  Discussion 12, 21 August 2006, Prague, Czech Republic, JD12, \#17}, 12,
  August 2006.

\bibitem{MWA}
Judd~D. Bowman et~al.
\newblock Field deployment of prototype antenna tiles for the {M}ileura
  {W}idefield {A}rray --- low frequency demonstrator.
\newblock {\em Astron. J.}, 133:1505--1518, 2007.

\bibitem{LOFAR}
H.~R. Butcher.
\newblock {L}{O}{F}{A}{R}: First of a new generation of radio telescopes.
\newblock {\em Proc. SPIE}, 5489:537--544, 2004.

\bibitem{ETA1}
S.~W. Ellingson, J.~H. Simonetti, and C.~D. Patterson.
\newblock Design and evaluation of an active antenna for a 29-47 {MH}z radio
  telescope array.
\newblock {\em IEEE Trans. Antenn. Propag.}, 55:826--831, 2007.

\bibitem{ETA2}
J.~H. Simonetti, S.~W. Ellingson, C.~D. Patterson, W.~Taylor, V.~Venugopal,
  S.~Cutchin, and Z.~Boor.
\newblock The {E}ight-meter-wavelength {T}ransient {A}rray ({ETA}).
\newblock {\em Bull. Amer. Astron. Soc.}, 37:1438--1438, 2006.

\bibitem{McLaughlin}
M.~A. McLaughlin, A.~G. Lyne, D.~R. Lorimer, M.~Kramer, R.~N. Faulkner, A.~J.
  Manchester, J.~M. Cordes, F.~Camilo, A.~Possenti, I.~H. Stairs, G.~Hobbs,
  N.~D'Amico, M.~Burgay, and J.~T. O'Brien.
\newblock Transient radio bursts from rotating neutron stars.
\newblock {\em Nature}, 439:817--820, 2006.

\bibitem{Lorimer}
D.~R. Lorimer, M.~Bailes, M.~A. McLaughlin, D.~J. Narkevic, and F.~Crawford.
\newblock A bright millisecond radio burst of extragalactic origin.
\newblock {\em Science}, 318:777--780, 2007.

\bibitem{Vachaspati}
Tanmay Vachaspati.
\newblock {Cosmic Sparks from Superconducting Strings}.
\newblock arXiv:0802.0711 [astro-ph], 2008.

\bibitem{UED}
Thomas Appelquist, Hsin-Chia Cheng, and Bogdan~A. Dobrescu.
\newblock Bounds on universal extra dimensions.
\newblock {\em Phys. Rev.}, D64:035002, 2001.

\bibitem{PDG}
W.~M. Yao et~al.
\newblock Review of particle physics.
\newblock {\em J. Phys.}, G33:1--1232, 2006.

\bibitem{RS1}
Lisa Randall and Raman Sundrum.
\newblock A large mass hierarchy from a small extra dimension.
\newblock {\em Phys. Rev. Lett.}, 83:3370--3373, 1999.

\bibitem{RS2}
Lisa Randall and Raman Sundrum.
\newblock {An alternative to compactification}.
\newblock {\em Phys. Rev. Lett.}, 83:4690--4693, 1999.

\bibitem{champ}
A.~Chamblin, S.~W. Hawking, and H.~S. Reall.
\newblock {Brane-world black holes}.
\newblock {\em Phys. Rev.}, D61:065007, 2000.

\bibitem{emp}
Roberto Emparan, Gary~T. Horowitz, and Robert~C. Myers.
\newblock {Exact description of black holes on branes}.
\newblock {\em JHEP}, 01:007, 2000.

\bibitem{CIN}
D.~Karasik, C.~Sahabandu, P.~Suranyi, and L.~C.~R. Wijewardhana.
\newblock Small (1-tev) black holes in randall-sundrum i scenario.
\newblock {\em Phys. Rev.}, D69:064022, 2004.

\bibitem{ADD}
Nima Arkani-Hamed, Savas Dimopoulos, and G.~R. Dvali.
\newblock The hierarchy problem and new dimensions at a millimeter.
\newblock {\em Phys. Lett.}, B429:263--272, 1998.

\bibitem{Emparan:2002jp}
Roberto Emparan, Juan Garcia-Bellido, and Nemanja Kaloper.
\newblock {Black hole astrophysics in AdS braneworlds}.
\newblock {\em JHEP}, 01:079, 2003.

\bibitem{Emparan:2002px}
Roberto Emparan, Alessandro Fabbri, and Nemanja Kaloper.
\newblock {Quantum black holes as holograms in AdS braneworlds}.
\newblock {\em JHEP}, 08:043, 2002.

\end{thebibliography}

\end{document}